\documentclass{elsart}
\usepackage{amsmath}
\usepackage{wasysym}
\usepackage{fancyhdr}
\usepackage{fullpage}
\usepackage{amsbsy}
\usepackage{amssymb}
\usepackage{amscd}
\usepackage{amsfonts}
\usepackage{supertabular}
\usepackage{graphics}
\usepackage{verbatim}
\usepackage{subfigure}
\usepackage{epsfig}
\usepackage{xspace}
\usepackage{euscript}
\usepackage{alltt}
\usepackage{boxedminipage}
\usepackage{float}
\usepackage[colorlinks]{hyperref}
\usepackage{color}
\usepackage[all]{xy}
\usepackage[authoryear]{natbib}
\usepackage{t1enc}
\usepackage{times,exscale}
\usepackage{graphicx,calc}
\usepackage{pstricks,psfrag}
\usepackage{psfrag}
\usepackage{color}
\usepackage{subfigure}

\def\br{{\mathbf{r}}}

\def\bR{{\mathbf{R}}}

\begin{document}
\pagestyle{fancyplain}

    \lhead[\fancyplain{}{\sl Radhakrishnan \& Gavini}]
          {\fancyplain{}{\sl Radhakrishnan \& Gavini}}
    \rhead[\fancyplain{}
    {\sl \hfill }]
    {\fancyplain{}
    {\sl \hfill }}

\begin{frontmatter}
\title{Orbital-free DFT study of the energetics of vacancy clustering and prismatic dislocation loop nucleation in aluminum}
\author{Balachandran Radhakrishnan,}
\author{Vikram Gavini\corauthref{cor}}
\address{Department of Mechanical Engineering, University of Michigan, Ann Arbor, MI 48109, USA}
\corauth[cor]{Corresponding Author (\it vikramg@umich.edu)}
\begin{abstract}
In the present work, we conduct large-scale orbital-free DFT calculations to study the energetics of vacancy clustering in aluminum from electronic structure calculations. The simulation domains considered in this study are as large as those containing a million atoms to accurately account for both the electronic structure and long-ranged elastic fields. Our results indicate that vacancy clustering is an energetically favorable mechanisms with positive binding energies for a range of vacancy clusters considered in the present study. In particular, the $19$ vacancy hexagonal cluster lying in $\{111\}$ plane has a very large binding energy with the relaxed atomic structure representative of a prismatic dislocation loop. This suggests that vacancy prismatic loops as small as those formed from 19 vacancies are stable, thus providing insights into the nucleation sizes of these defects in aluminum.
\end{abstract}


\end{frontmatter}

\section{Introduction}
Prismatic dislocation loops play an important role in influencing the macroscopic mechanical properties of materials, in particular ductility and fracture toughness in metals~\citep{Suresh2000, Lubarta2004, Wirth2007}. A large concentration of prismatic loops have been observed in quenched metals and in materials subject to large doses of radiation. It is widely believed that vacancy clustering is a precursor mechanism to the nucleation of prismatic loops in quenched metals~\citep{kuhlmann1960behavior, Cotterill1963}. In irradiated materials, experimental studies have shown a direct correlation between the dose of irradiation, the population of prismatic dislocation loops, and the loss of ductility and fracture toughness (cf. e.g.~\citet{Barnes1960,eyre1965electron,Masters1965,trinkaus1997segregation,singh1997radiation,Zinkle1998,Zinkle2011}). Numerous atomistic simulations have been conducted using empirical interatomic potentials to study the nucleation and evolution of defects in irradiated materials (cf. e.g.~\citet{Bacon1993,robinson1994basic,Bacon1994,Ackland1997,Rubia1998,trachenko2001atomistic,Wirth2000,han2003interatomic,Marian2002PRL,Marian2002PRB,Caturla2006}). These studies revealed displacement cascades nucleating a large concentration of vacancies and self-interstitals, which subsequently result in the formation of prismatic dislocation loops (vacancy loops and self-interstital clusters), among other defects, through the coalescence of vacancies into vacancy clusters and self-interstitials into self-interstitial clusters. 

While the displacement cascade simulations have elucidated the overall mechanism of the formation of prismatic loops, they do not provide insights into the energetics of formation of these defects which dictates the nucleation and evolution of these defects. Further, the nucleation of prismatic loops is a very rapid process that is hard to capture experimentally, especially given the high mobility of nanometer-sized prismatic loops~\citep{Arakawa2007,Zinkle2007}. To this end, atomistic calculations have been employed to study the energetics of various sizes of vacancy and self-interstitial clusters (cf. e.g.~\citet{Wirth1997,Wirth2000,Marian2002PRL,Morishita2003}). While these atomistic studies have provided many important qualitative insights, these efforts have used empirical potentials to model interactions between various atoms in materials. The results of these simulations are significantly influenced by the choice of empirical potentials used and the physical quantities to which the parameters of these empirical potentials are fitted. It is hard to ascertain the accuracy of these empirical potentials to describe the details of the defect core, which involves situations with making and breaking of chemical bonds governed by quantum-mechanical interactions. Thus, it is desirable to conduct density functional theory (DFT) calculations to study the energetics of the formation of these defects. The most widely used implementations of density functional theory are based on Fourier space formulations with a plane-wave discretization. While these Fourier space implementations have provided tremendous insights into the bulk properties of a wide range of materials, they are often restricted to simulation domains containing a few hundred atoms with periodic geometries. This poses a severe restriction in the electronic structure study of defects which require non-periodic geometries and larger simulation domains to accurately account for both the quantum-mechanical interactions at the defect core as well as the long-ranged elastic fields.   

In the present work we employ a real-space formulation of orbital-free density functional theory and the quasi-continuum reduction technique to conduct electronic structure calculations on multi-million atom systems. In particular, we employ the Wang Govind Carter (WGC) orbital-free kinetic energy functional~\citep{Yan2} which has been shown to be accurate for aluminum for a wide range of properties~\citep{Carter-Al-Mg, Das2015}. We begin our study by computing the binding energies of divacancies in aluminum. A cell-size study of the binding energy of divacancies has shown significant cell-size dependence up to 2,000 atoms, underscoring the need for large-scale electronic structure calculations to accurately study the energetics of defects in materials. The computed binding energies of both $\left<100\right>$ and $\left<110\right>$ divacancies are positive indicating the tendency of vacancies to attract. To understand the energetics of vacancy clustering, we next consider quad-vacancy clusters formed from divacancies. The computed binding energies of all quad-vacancy clusters considered in this study are positive. Among the planar quad-vacancy clusters, those lying in the \{111\} plane had the highest binding energy. In order to investigate the energetics of vacancy clusters in the \{111\} plane, which is also one of the habit planes for vacancy prismatic loops, we consider hexagonal vacancy clusters of various sizes on this plane. While the 7 vacancy hexagonal cluster has positive binding energy, this is only marginally greater than the quad-vacancy cluster in the \{111\} plane. However, the 19 vacancy cluster has a very large relaxed binding energy and the atomic structure closely resembles a collapsed prismatic dislocation loop. This study suggests that vacancy clusters containing as small as 19 vacancies can collapse to form stable prismatic loops, thus providing insights into the nucleation size of these defects.  

The remainder of this paper is organized as follows. Section~\ref{sec:method} presents a brief overview of the local real-space formulation of orbital-free DFT and the quasi continuum reduction technique employed in this work that has enabled consideration of multi-million atom computational domains. Section~\ref{sec:Results} presents our electronic structure study of vacancy clustering and nucleation of prismatic dislocation loops, along with a discussion of the new findings and their implications. We finally conclude with an outlook in Section~\ref{sec:Conclusion}.

\section{Overview of Methodology}\label{sec:method}
In this section, for the sake of completeness and to keep the discussion self-contained, we provide an overview of the local real-space formulation of orbital-free DFT, finite-element discretization, and the coarse-graining technique---quasi-continuum orbital-free DFT---employed in this work. We refer to~\citet{GaviniPRB2010,Motamarri-OFDFT2012} for a comprehensive discussion on the local real-space formulation, and~\citet{Gavini2, GaviniPRB2010} for details on the coarse-graining technique. The main ideas are discussed below.

\subsection{Local real-space formulation of orbital-free DFT}\label{sec:RS-OFDFT}
The ground-state energy in orbital-free density functional theory~\citep{Parr} is given by
\begin{equation}\label{eqn:Energy}
E(u,\bR) = T_{s}\left(u\right) + E_{xc}\left(u\right) + E_{H}\left(u\right) + E_{ext}\left(u,\bR\right) + E_{zz}\left(\bR\right)\,,
\end{equation}
where, $u$ denotes the square-root of electron-density and $\bR=\{\bR_1,\bR_2,\ldots,\bR_M\}$ denotes the vector collecting the positions of atoms. In equation~\eqref{eqn:Energy}, $T_{s}$ denotes the kinetic energy of non-interacting electrons, which is explicitly modeled in orbital-free DFT; $E_{xc}$ denotes the exchange-correlation energy, which includes all the quantum-mechanical interactions between electrons; $E_H$ denotes the Hartree energy or the classical electrostatic interactions between the electrons; $E_{ext}$ denotes the classical electrostatic interaction energy between the electrons and nuclei; and $E_{zz}$ denotes the nuclear-nuclear electrostatic repulsion energy. 

The density-dependent Wang-Govind-Carter (WGC) kinetic energy functional~\citep{Yan2} is the most widely used model for $T_{s}$ in solid-state orbital-free DFT calculations on materials systems whose electronic structure is close to a free-electron gas, and is employed in this work. Numerical studies have indicated that this is a transferable model for Al, Mg and Al-Mg materials systems with accuracies commensurate with Kohn-Sham DFT calculations on a wide range of materials properties~\citep{Carter-Al-Mg,Ho,Das2015}. The functional form of the WGC kinetic energy functional is given by
\begin{equation}\label{eqn:KE}
T_{s}\left(u\right)=C_{F}\int {u^{{10}/{3}} d\br} + \frac{1}{2}\int {\left|\nabla u\left(\br \right)\right|^{2} d\br} + T^{\alpha,\beta}_{\rm K} (u) \,,
\end{equation}
where
\begin{equation}
T^{\alpha,\beta}_{\rm K}(u)=C_F\int\int u^{2\alpha}(\br) K(|\br-\br'|; u(\br),u(\br')) u^{2\beta}(\br') d\br d\br' \,\,.
\end{equation}
In the above equation, the first term is the Thomas-Fermi energy with $C_F=\frac{3}{10}(3\pi^2)^{2/3}$, the second term is the von-Weizs$\ddot{a}$cker correction~\citep{Parr}, and the last term, $T^{\alpha,\beta}_{\rm K}$, denotes the density-dependent kernel kinetic energy functional. The kernel $K$ and parameters $\alpha$ and $\beta$ are chosen such that the linear response of uniform electron gas matches the theoretically known Lindhard response~\citep{Finnis}. In particular, in the WGC functional, these parameters are chosen to be $\{\alpha,\beta\}=\{5/6+\sqrt{5}/6,5/6-\sqrt{5}/6\}$, and the density-dependent kernel is expanded as a Taylor series about the bulk average electron density, resulting in a series of density independent kernels~\citep{Yan2}.

Widely used models for the exchange-correlation energy, especially for solid-state calculations of ground-state properties, include the local density approximation (LDA) and the generalized gradient approximation (GGA) (cf.~e.g.~\citet{Martin}), and have been demonstrated to be transferable models for a range of materials systems and materials properties. In the present work, we will employ the LDA exchange-correlation functional given by
\begin{equation}\label{exc}
E_{xc}(u) = \int \varepsilon_{xc}(u)u^{2}(\br) d\br \,,
\end{equation}
where $\varepsilon_{xc} = \varepsilon_{x} + \varepsilon_{c}$ is the exchange and correlation energy
per electron given by
\begin{equation}
\varepsilon_x(u) = -\frac{3}{4}\left(\frac{3}{\pi}\right)^{1/3}u^{2/3},
\end{equation}
\begin{equation}
\varepsilon_c(u) = \begin{cases}
&\frac{\gamma}{(1 + \beta_1\sqrt(r_s) + \beta_2r_s)}\;\;\;\;\;\;\;\;\;\;\;\;\;\;\;\;\;\;\;\;\;\;\;r_s\geq1\\
&A\,\log r_s + B + C\,r_s\log r_s + D\,r_s\;\;\;\;\;\;\;\;r_s\,<\,1,
\end{cases}
\end{equation}
where $r_s = (\frac{3}{4\pi u^2})^{1/3}$. The values of constants used in this study are those of an unpolarized medium, and are given by $\gamma_u$ = -0.1471, $\beta_{1u}$ = 1.1581, $\beta_{2u}$ = 0.3446, $A_{u}$ = 0.0311, $B_u$ = -0.048, $C_u$ = 0.0014, $D_u$ = -0.0108.

The remaining terms in equation~\eqref{eqn:Energy} constitute the classical electrostatic energy between electrons and the nuclei, and are given by
\begin{align}
E_{H}(u) &= \frac{1}{2}\int\int\frac{u^2(\br)u^2(\br')}{|\br - \br'|}d\br d\br' \,,\label{hartree}\\
E_{ext}(u,\bR) &= \int u^2(\br) V_{ext}(\br) d\br = \sum_J\int u^2(\br) V^J_{ps}(\br,\bR_J) d\br \,,\label{external}\\
E_{zz} &= \frac{1}{2}\sum_{I,J \neq I} \frac{Z_I Z_J}{|\bR_I-\bR_J|}\,, \label{repulsive}
\end{align}
where $Z_{J}$ and $V^J_{ps}(\br,\bR_J)$ denote the valence charge and pseudopotential corresponding to atom $J$ located at $\bR_{J}$, respectively.

In the energy functional~\eqref{eqn:Energy}, all the terms are local, excepting the extended interactions in electrostatic energy and the kernel energy. A local variational reformulation of these extended interactions in real-space has been developed in prior works~\citep{Gavini1,GaviniPRB2010}, which has enabled the consideration of general boundary conditions for studies on energetics of isolated defects~\citep{Iyer2015,Das2015} and is also a crucial step in developing coarse-graining schemes for electronic structure calculations. The extended interactions in electrostatics are governed by the $\frac{1}{|\br-\br'|}$ kernel, which is the Green's function of the Laplace operator. Thus, the total electrostatic interaction energy can be reformulated as the following local variational problem:
\begin{equation}\label{eqn:elReformulation}
E_H+E_{ext}+E_{zz}
= -\inf_{\phi \in \mathcal{Y}} \left\{\frac{1}{8\pi}\int |\nabla \phi(\br)|^2 d\br - \int \big(u^{2}(\br) + \sum_{I=1}^{M}b_{I}(\br,\bR_I)\big)\phi(\br)d\br\right\}-E_{self}\,.
\end{equation}
In the above, $b_I(\br,\bR_I)$ denotes the nuclear charge distribution corresponding to the ionic pseudopotential of the $I^{th}$ nucleus, $\phi$ denotes the electrostatic potential corresponding to the total charge distribution comprising of the electrons and the nuclei, and $\mathcal{Y}$ is an appropriate function space. $E_{self}$ denotes the self energy of the nuclear charge distributions which is computed by taking recourse to the Poisson equation associated with nuclear charge distribution (cf.~\citet{Motamarri-OFDFT2012, Motamarri-KSDFT2013}).

\citet{Choly2002} proposed an approach to develop a local real-space formulation for the extended interactions in the kernel kinetic energy functional. In particular, they demonstrated that the series of density independent kernels obtained from the Taylor series expansion of the WGC density dependent kernel can each be approximated in the Fourier-space using a sum of partial fractions. Using this approximation, the extended interactions in the kernel kinetic energy functional can be reformulated in terms of the solutions of a series of Helmholtz equations.  If $\bar{K}(|\br-\br'|)$ denotes a density independent kernel, and is approximated in the Fourier-space using the approximation $\widehat{\bar{K}}(\mathbf{q})\approx \sum_{j=1}^{m}\frac{A_j |\mathbf{q}|^2}{|\mathbf{q}|^2+B_j}$, the kernel energy corresponding to the density independent kernel can be expressed as the following variational problem~\citep{GaviniPRB2010,Motamarri-OFDFT2012}:
\begin{equation}\label{eqn:kerReformulation}
\int\int u^{2\alpha}(\br) \bar{K}(|\br-\br'|) u^{2\beta}(\br') d\br d\br'= \inf_{{\raisebox{-0.7ex}{$\scriptstyle\mathbf{\tilde{\omega}}_{\alpha}\in \mathcal{Z}$}}}\,\sup_{\mathrm{\tilde{\omega}}_{\beta} \in \mathcal{Z}}\bar{L}(u,\mathbf{\tilde{\omega}}_{\alpha},\mathbf{\tilde{\omega}}_{\beta})\,,
\end{equation}
where
\begin{equation}
\bar{L}(u,\mathbf{\tilde{\omega}_{\alpha}},\mathbf{\tilde{\omega}_{\beta}}) =  \sum_{j=1}^{m}\Bigl\{\int \Bigl[\frac{1}{A_j\;B_j}\nabla\omega_{\alpha_j}\cdot\nabla\omega_{\beta_j} + \frac{1}{A_j}\omega_{\alpha_j}\omega_{\beta_j} + \omega_{\beta_j}u^{2\alpha} + \omega_{\alpha_j}u^{2\beta} + A_j u^{2(\alpha+\beta)}\Bigr]d\br\Bigr\}\,.
\end{equation}
In the above, $\mathbf{\tilde{\omega}}_{\alpha}=\{\omega_{\alpha_1},\omega_{\alpha_2},\ldots,\omega_{\alpha_m}\}$ and $\mathbf{\tilde{\omega}}_{\beta}=\{\omega_{\beta_1},\omega_{\beta_2},\ldots,\omega_{\beta_m}\}$ denote the vector of potential fields, and $\mathcal{Z}$ denotes a suitable function space. We refer by `kernel potentials' the auxiliary potential fields, $\omega_{\alpha_j}$ and $\omega_{\beta_j}$ for $j=1\ldots m$, introduced in the local reformulation of the extended interactions in the kernel energy.

The problem of computing the ground-state energy and ground-state electronic structure of an $N_{e}$ electron system in orbital-free DFT for a fixed position of atoms can be expressed as a local variational problem in real-space using the reformulations in equations~\eqref{eqn:elReformulation} and ~\eqref{eqn:kerReformulation}, and is given by 
\begin{equation}
\inf_{{\raisebox{-0.7ex}{$\scriptstyle u\in \mathcal{X}$}}} \, \sup_{\phi \in \mathcal{Y}} \,\, \inf_{{\raisebox{-0.7ex}{$\scriptstyle\mathbf{\tilde{\omega}}_{\alpha}\in \mathcal{Z}$}}}\,\sup_{\mathrm{\tilde{\omega}}_{\beta} \in \mathcal{Z}} L(u,\phi,\mathbf{\tilde{\omega}}_{\alpha}, \mathbf{\tilde{\omega}}_{\beta}) \quad\qquad \mbox{subject to:} \int u^2(\br) d\br = N_e \,,
\end{equation}
where $L$ denotes the Lagrangian resulting from the local reformulations in equations ~\eqref{eqn:elReformulation} and ~\eqref{eqn:kerReformulation}. The function spaces, $\mathcal{X}$, $\mathcal{Y}$ and $\mathcal{Z}$ are appropriately chosen based on the boundary conditions dictated by the problem. The well-posedness of this local variational saddle-point orbital-free DFT problem has been established using the direct method in calculus of variations for certain models of orbital-free DFT kinetic energy functionals, and we refer to~\citet{Gavini1} for further details on the mathematical properties of the formulation.

\subsection{Finite-element discretization}\label{sec:OFDFT-FE}
A finite-element basis presents a natural basis set to discretize the local real-space variational formulation of orbital-free DFT discussed in section~\ref{sec:RS-OFDFT}, and is employed in this work. The finite-element discretization of the orbital-free DFT problem presents many advantages over the widely used plane-wave discretization of orbital-free DFT calculations~\citep{Profess2}. It allows for the consideration of complex geometries and boundary conditions that are not accessible through Fourier-space formulations of orbital-free DFT employing plane-wave discretization. This freedom from periodic boundary conditions is significant in the study of the energetics of defects in materials, which often break periodicity of perfect materials. To elucidate, the geometry of an isolated dislocation is not compatible with periodic boundary conditions which has limited our capabilities in studying the energetics of such defects. The recent developments in real-space formulation of orbital-free DFT and the finite-element discretization have enabled some of the first studies on the energetics of an isolated edge dislocation in aluminum, providing key insights into the size of a dislocation-core or core-energetics from electronic structure calculations~\citep{Iyer2015}. While Fourier space implementations are computationally more efficient than real-space implementations using finite-element basis, recent numerical efforts which use higher-order finite-element discretizations~\citep{Motamarri-OFDFT2012} have been shown to bridge this gap. Further, the good scalability of the finite-element discretization on parallel computing platforms allows the use of high performance computing resources to consider complex problems that are not accessible otherwise. Finally, the finite-element discretization is amenable to unstructured coarse-graining, which is the main idea exploited in the quasi-continuum coarse-graining technique, discussed subsequently, which has enabled electronic structure calculations at macroscopic scales using orbital-free DFT.  

\subsection{Quasi-continuum orbital-free density functional theory}\label{sec:QCOFDFT}
In this section, we present the key ideas in quasi-continuum orbital-free DFT (QC-OFDFT)---a seamless coarse-graining approach that enables us to perform multi-million atom orbital-free DFT simulations for electronic structure studies on defects at physically realistic concentrations. QC-OFDFT uses orbital-free DFT as the sole input physics and exploits the local real-space formulation in conjunction with the basis set adaptivity of the finite-element basis to achieve orbital-free DFT electronic structure calculations at macroscopic scales. The quasi-continuum reduction technique for orbital-free DFT was first demonstrated for local kinetic energy functionals~\citep{Gavini2} and later extended to the non-local WGC kinetic energy functionals employed in this work~\citep{GaviniPRB2010}.\par

The idea of quasi-continuum reduction was originally proposed in the context of interatomic potentials as a seamless scheme bridging the atomistic and continuum length-scales for crystalline materials~\citep{Tadmor1996,knap2001analysis} in order to simultaneously account for the atomistic interactions at the defect-core as well as the long-ranged elastic fields accompanying the defect. The quasi-continuum reduction idea is based on kinematic constraints introduced of the positions of atoms, thus reducing the number of degrees of freedom in the variational problem of computing the ground-state of a materials system. The kinematic constraints are introduced via a finite-element triangulation of a subset of atoms in the computational domain, which are referred to as the representative atoms or \emph{rep atoms}. The \emph{rep atoms} are chosen such that full atomistic resolution is provided in regions of interest, such as the defect-core with large atomistic displacements, and coarse-graining elsewhere. The kinematic constraints on atomistic displacements introduced through the finite-element triangulation of the \emph{rep atoms} provides an excellent subspace for solving the variational problem of computing the ground-state energy of crystalline materials systems with defects.\par

\begin{figure}[htbp]
\begin{center}
\includegraphics[width=1.0\textwidth]{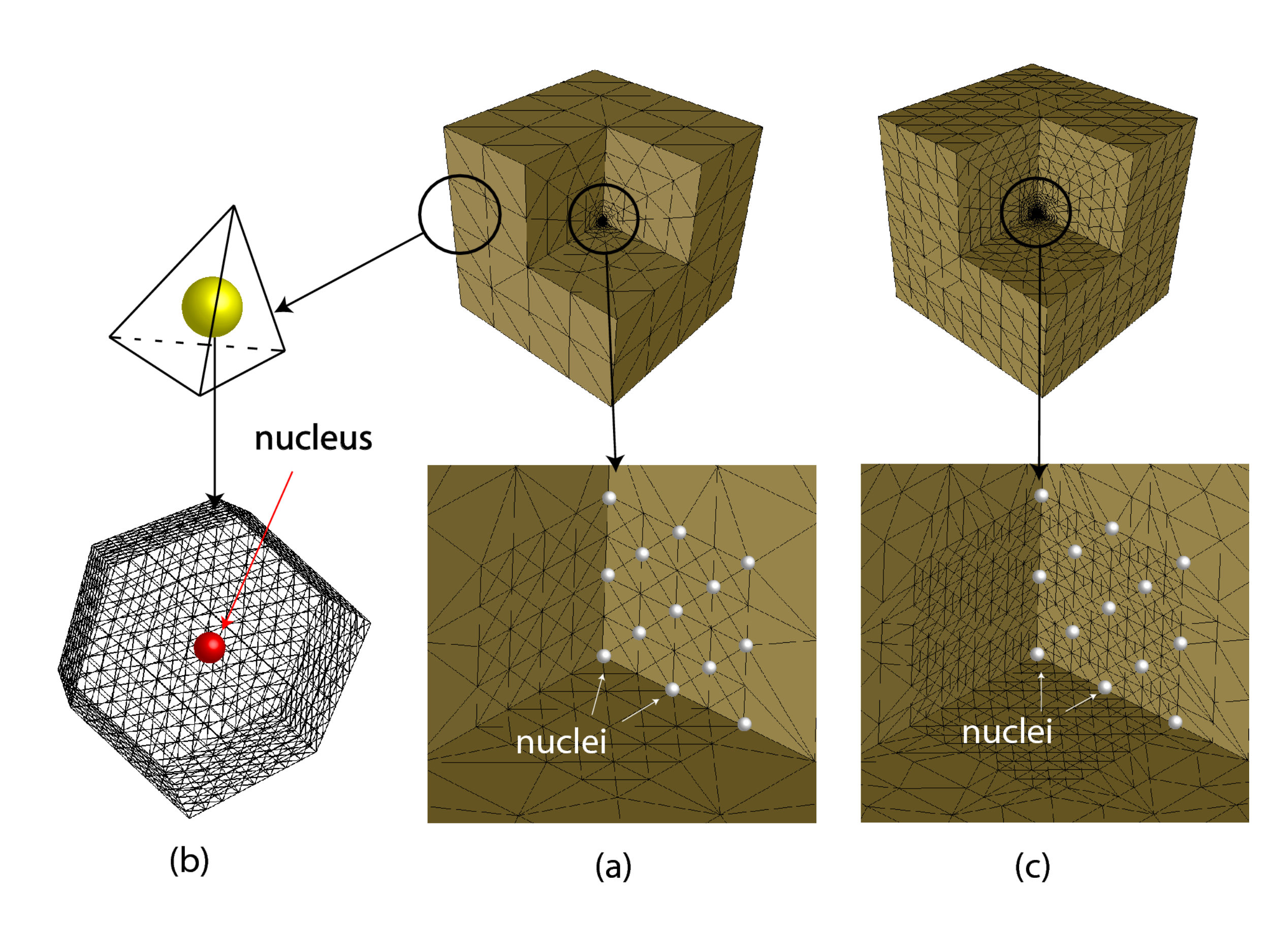}
\caption{\label{fig:qcMesh} (a) Atomic mesh used to represent positions of nuclei. (b) Auxiliary unit-cell meshes used to represent the predictor fields. (c) Electronic-mesh used to represent the corrector fields, which has subatomic resolution in the defect-core and coarsens away from the defect-core becoming superatomic.}
\end{center}
\end{figure}

The quasi-continuum reduction for electronic structure calculations presents an additional challenge of coarse-graining the electronic fields that exhibit oscillations on a subatomic length-scale. In QC-OFDFT formalism, the atomic displacements are treated in a similar manner as the quasi-continuum approach for interatomic potentials where kinematic constraints are introduced on atomistic displacements using a finite-element triangulation of \emph{rep atoms}. The electronic fields comprising of the square-root electron-density, electrostatic potential, and kernel potentials are decomposed into predictor and corrector fields:  
\begin{equation}
\begin{split}
u= & u_{0} + u_{c}\,,\\
\phi= & \phi_{0} + \phi_{c}\,, \\
\omega_{\alpha}= & {\omega_{\alpha}}_{0} + {\omega_{\alpha}}_{c}\,, \\
\omega_{\beta}= & {\omega_{\beta}}_{0} + {\omega_{\beta}}_{c},
\end{split}
\label{eqn:decomp}
\end{equation}
where ($u_{0}$, $\phi_{0}$, ${\omega_{\alpha}}_{0}$, ${\omega_{\beta}}_{0}$) denote the predictor electronic-fields that are computed using periodic unit-cell calculations undergoing the Cauchy-Born deformation of the underlying lattice. The predictor fields provide a good representation of the electronic fields to the leading order in regions of smooth deformations~\citep{Blanc}, such as regions away from the defect-core that are governed by elastic interactions. However, the electronic structure deviates significantly from that of the predictor fields in regions of rapidly varying deformations, such as in the defect-core. These deviations are captured through the corrector fields ($u_{c}$, $\phi_{c}$, ${\omega_{\alpha}}_{c}$, ${\omega_{\beta}}_{c}$), and are computed from the local variational saddle-point problem discussed in section~\ref{sec:RS-OFDFT}. As the predictor fields are a good representation of the electronic fields away from the defect-core, the corrector fields are significant only near the defect-core. Thus, the corrector fields can be effectively represented by a finite-element triangulation which has sufficient resolution near the defect-core, but coarse-grains far away. The numerical implementation of the QC-OFDFT method is conducted using a hierarchy of finite-element triangulations (cf. figure~\ref{fig:qcMesh}): (i) an atomic-mesh representing atomic displacements, which has full atomistic resolution in the vicinity of the defect-core and coarse-grains far away; (ii) electronic-mesh representing the corrector electronic fields, which has subatomic resolution in the vicinity of the defect-core and coarse-grains away becoming superatomic; (iii) auxiliary unit-cell meshes used to represent the predictor electronic fields. Often, for the sake of convenience, the electronic-mesh is chosen to be a subgrid of the atomic mesh. Using the decomposition in equation~\eqref{eqn:decomp}, the problem of computing the ground-state electronic structure reduces to a saddle-point variational problem in the coarse-grained independent variables comprising of the corrector fields on the electronic-mesh. We refer to~\citet{Gavini2} for further details and a comprehensive discussion of the method.    

Numerical studies have shown that the QC-OFDFT method converges rapidly with respect to coarse-graining, where a few thousand \emph{rep atoms} have been sufficient to represent computational domains nominally containing millions of atoms for studying energetics of point defects. The effectiveness of QC-OFDFT has made possible electronic structure calculations at macroscopic scales using orbital-free DFT for studying the energetics of defects at realistic concentrations. Further, this provides a seamless scheme to account for both the quantum-mechanical effects at the defect-core and the long-ranged elastic fields using an electronic structure theory (orbital-free DFT) as the sole input physics. Prior investigations have used QC-OFDFT to study the energetics of vacancies and dislocations in aluminum, and have provided important physical insights~\citep{Gavini3,GaviniPRL2008,GaviniPRB2010}. We refer to~\citet{Iyer2011field,GaviniJMPS2011} for numerical and mathematical analysis of the quasi-continuum reduction of orbital-free DFT, and field theories, in general.   

\section{Results and Discussion}\label{sec:Results}
We now present the results of our study on the energetics of vacancy clustering in aluminum and nucleation of vacancy prismatic dislocation loops using large-scale orbital-free DFT electronic structure calculations. In this study, we employ the Wang-Govind-Carter (WGC) kinetic energy functional~\citep{Yan2} with first-order Taylor expansion of the density dependent kernel (cf.~\citet{Yan2}), a local density approximation (LDA) for the exchange-correlation energy~\citep{Perdew}, and bulk local pseudopotential for aluminum (BLPS)~\citep{zhou2004transferable}. The WGC kinetic energy functional and the BLPS pseudopotential have been shown to be accurate and transferable for a range of material properties of Al, Mg and Al-Mg materials systems~\citep{Das2015}. Numerical parameters such as finite-element discretization, coarsening of finite-element triangulation in the quasi-continuum method, numerical quadratures, and stopping tolerances on iterative non-linear and linear solvers are chosen such that the errors in the computed formation energies do not exceed $0.01$ $eV$. In all the simulations reported in this work, we employ homogeneous Dirichlet boundary conditions on the corrector electronic fields, which correspond to the perturbations in the electronic fields arising from the defect vanishing on the boundary of the simulation domain with the electronic structure beyond the computational domain corresponding to that of the bulk. We refer to these boundary conditions on electronic fields as bulk Dirichlet boundary conditions which simulate an isolated defect in a bulk crystalline material. We hold the positions of atoms fixed on the boundary of the simulation cell, while relaxing the positions of the interior atoms. 

\subsection{Divacancies} 
In order to understand cell-size effects and establish the simulation domain sizes needed to accurately understand the energetics of vacancy interactions, we first conduct a cell-size study on divacancy binding energies in aluminum. The binding energy of an $n$-vacancy cluster is computed as:
\begin{equation}
E^{bind}_{nv} =  nE^{f}_{v}-E^{f}_{nv}\,,
\label{eqn:Ebind}
\end{equation}
where $E^{f}_{nv}$ is the formation energy of the $n-$vacancy cluster and $E^{f}_{v}$ is the formation energy of the monovacancy. The formation energy (at constant volume) of an $n-$vacancy cluster is given by~\citep{Finnis}
\begin{equation}
E^{f}_{nv} =  E\left(N-n,n,\frac{N-n}{N}V\right) - \frac{N-n}{N}E\left(N,0,V\right),
\label{eqn:nvFormtion}
\end{equation}
where $E\left(N-n,n,\frac{N-n}{N}V\right)$ denotes the energy of the system with $N$ lattice sites occupied by the $n$-vacancy cluster and $N-n$ atoms with the total volume of the system being $\frac{N-n}{N}V$. In the above, $E\left(N,0,V\right)$ denotes the energy of an $N$-atom perfect crystal occupying a volume $V$. We compute the formation energies and binding energies at the equilibrium volume of a perfect crystal as we are primarily interested in the energetics of vacancy interactions in stress-free solids. We note that recent studies indicate that macroscopic deformations and macroscopic stresses influence the energetics of defects in a significant way~\citep{GaviniPRL2008,Iyer2014,Iyer2015}, but this is not the focus of the present study and will be a topic for future investigations. 	
\begin{figure}[htbp]
\centering
\scalebox{0.5}{\includegraphics{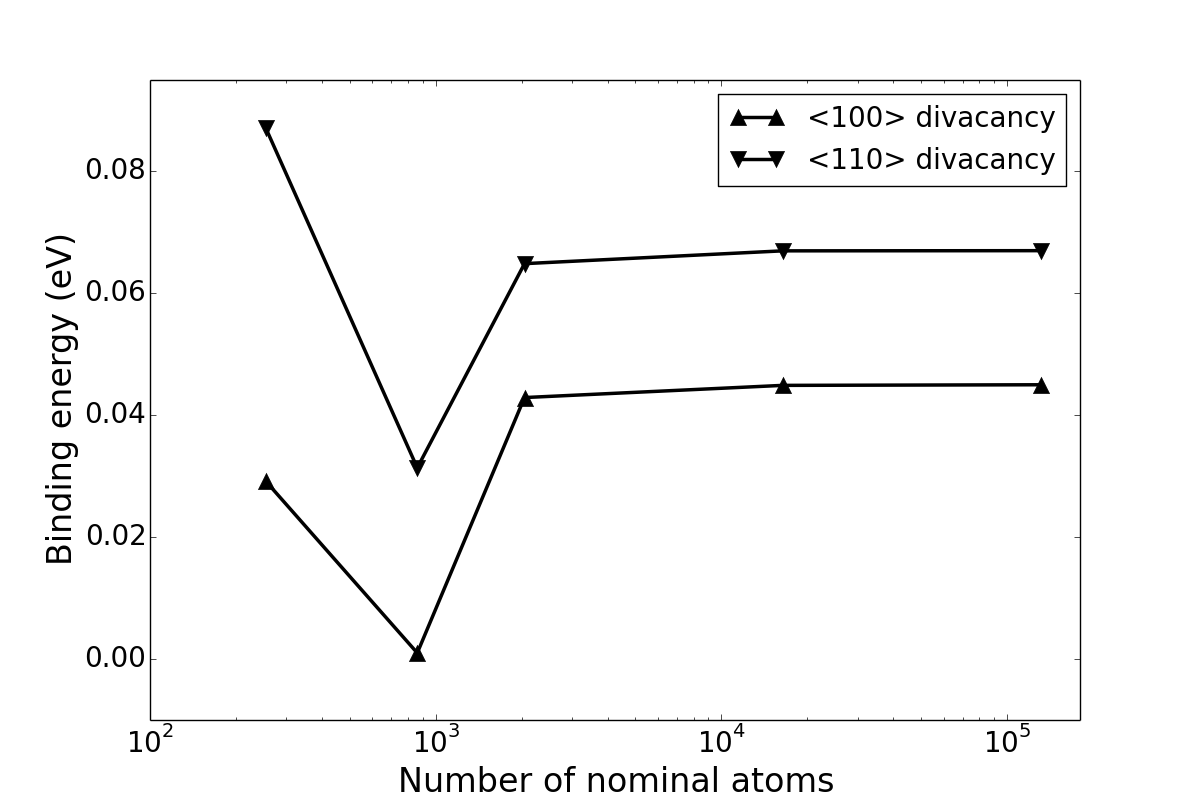}}
\caption{\label{fig:divacbinding} Cell-size study of binding energies of divacancies along $\left<100\right>$ and $\left<110\right>$ directions in aluminum.}
\end{figure}

Figure~\ref{fig:divacbinding} shows the computed binding energies of divacancies along $\left<100\right>$ and $\left<110\right>$ crystallographic directions for cell-sizes ranging from 32 lattice sites to over 100,000 lattice sites. The results suggest a significant cell-size dependence on the computed binding energies. In particular, this study indicates that cell-sizes containing about $2,000$ nominal number of atoms are required to obtain convergence in the binding energies of these simple defects. These results underscore the long-ranged nature of the electronic fields and elastic displacement fields in the presence of defects, and emphasize the need for coarse-graining techniques such as QC-OFDFT for studying the energetics of defects. We note that the observed cell-size effects are in agreement with prior QC-OFDFT cell-size studies on mono- and di-vacancies formation energies in aluminum~\citep{Gavini2,GaviniPRB2010}. The converged $\left<100\right>$ and $\left<110\right>$ divacancy relaxed binding energies are computed to be $0.045$ $eV$ and $0.067$ $eV$, respectively. The positive binding energies of the divacancies suggest that vacancies tend to attract and form a stable divacancy complex as opposed to remaining as separated monovacancies. 

\subsection{Quad-vacancies}
Having established the stability of divacancies, we next proceed to further investigate the vacancy clustering mechanism by computing the binding energies of quad-vacancy clusters formed from divacancies. As the number of possible quad-vacancy clusters is very large, we restrict our study to those quad-vacancy configurations where each vacancy has two other vacancies as the nearest or second nearest neighbors. This results in nine configurations, six of which are planar quad-vacancy clusters and three are non-planar configurations. Table~\ref{tab:quadVacBind} lists these configurations and the computed relaxed binding energies. In all these simulations, informed from the cell-size study of divacancies, we use cell-sizes containing $16,384$ nominal number of atoms to ensure convergence with respect to cell-size.
	
\begin{table}[h]
\caption{Binding energy of quad-vacancies in aluminum}
\centering 
\scalebox{1.0}{
\begin{tabular} {c  c  c}
\hline
Structure & Vacancy positions & Binding energy (eV) \\ \hline
Planar \{100\} & (0,0,0), (a/2,a/2,0), (a,0,0), (a/2,-a/2,0)& 0.08 \\
Planar \{100\} &(0,0,0), (a/2,a/2,0), (a,0,0), (3a/2,a/2,0) & 0.1\\
Planar \{100\} &(0,0,0), (a/2,a/2,0), (a,0,0), (a,a,0) & 0.1\\
Planar \{100\} & (0,0,0), (a,0,0), (0,a,0), (a,a,0) & 0.22\\
Planar \{110\} & (0,0,0), (0,a/2,a/2), (a,0,0), (a,a/2,a/2)& 0.3\\
Planar \{111\} & (0,0,0), (0,a/2,a/2), (a/2,a/2,0), (a/2,a,a/2)& 0.33\\
Non Planar &(0,0,0), (0,a/2,a/2), (a/2,0,a/2), (a/2,a/2,0) & 0.55\\
Non Planar & (0,0,0), (a,0,0), (a/2,a/2,0), (a/2,0,a/2)& 0.31\\
Non Planar &(0,0,0), (a,0,0), (a/2,a/2,0), (0,a/2,a/2) & 0.23\\ \hline
\end{tabular}
}
\label{tab:quadVacBind}
\end{table}
	
\begin{figure}[htbp]%
\centering%
\scalebox{1.2}{\includegraphics{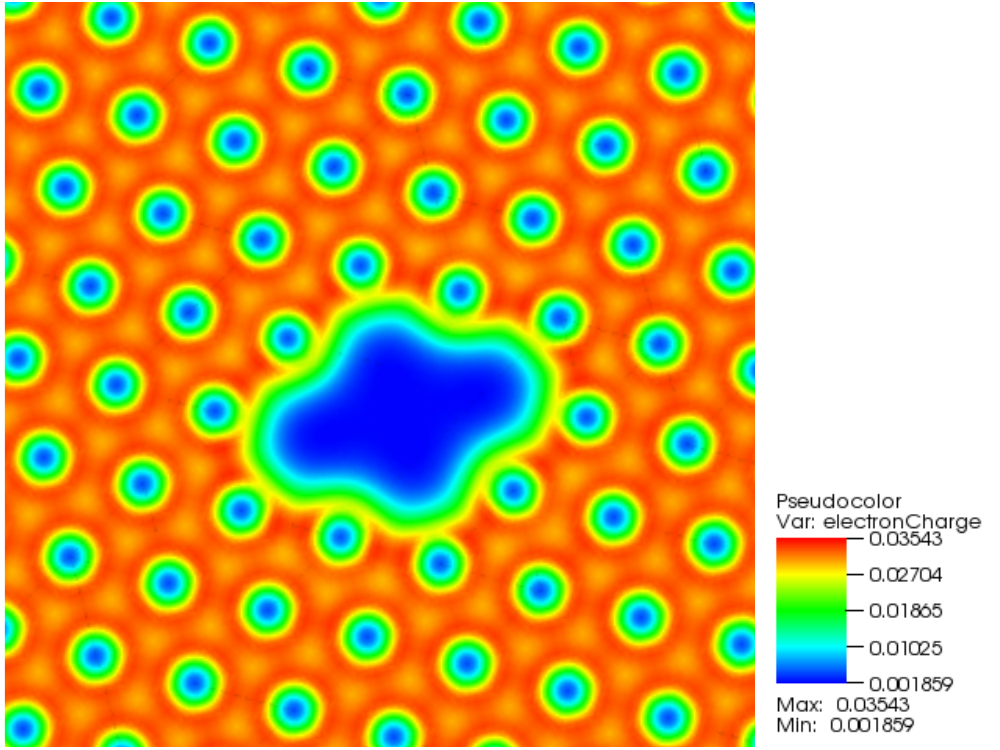}}%
\caption{\label{fig:contour111} Electron-density contours of quad-vacancy cluster lying in $\{111\}$ plane.}%
\end{figure}
	
\begin{figure}[htbp]%
\centering%
\scalebox{1.2}{\includegraphics{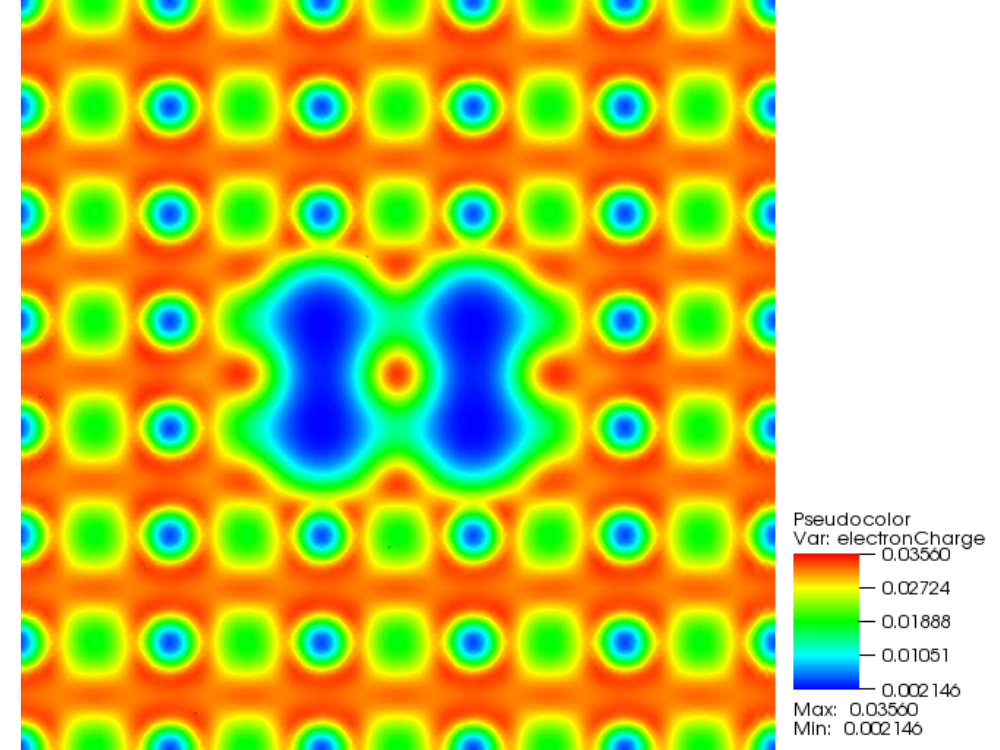}}%
\caption{\label{fig:contour110} Electron-density contours of quad-vacancy cluster lying in $\{110\}$ plane.}%
\end{figure}

The computed binding energies for all quad-vacancy clusters are positive suggesting an energetic preference to remain as quad-vacancy clusters as opposed to dissociation into monovacancies. However, the binding energies of the first three planar configurations of quad-vacancies lying in the \{100\} planes in table~\ref{tab:quadVacBind} are less than twice the binding energy of $\left<110\right>$ divacancy, suggesting that these quad-vacancy clusters are likely to dissociate into a pair of divacancies. On the other hand, the binding energies of other planar and non-planar quad-vacancy clusters are much larger and are stable with respect to dissociation into divacancies. Among the planar quad-vacancy clusters, those lying in the $\{111\}$ and  $\{110\}$ planes are the most stable. Figures \ref{fig:contour111} and \ref{fig:contour110} show the electron-density contours of quad vacancies lying in the $\{111\}$ and $\{110\}$ planes, respectively. Experimental studies suggest that these planes are the most likely habit-planes for vacancy clustering, eventually resulting in the formation of dislocation loops~\citep{kuhlmann1960behavior,Cotterill1963,Wu2007}. The non-planar configurations are tetrahedral quad-vacancy clusters, with the largest binding energy corresponding to a stacking fault tetrahedra. Stacking fault tetrahedra formed from larger vacancy clusters play an important role in influencing plastic deformation in materials~\citep{Kiritani1997,Kiritani1999}, and a study of the energetics of these defects will be a topic for future studies. In the present work, we will restrict ourselves to the study of larger planar vacancy clusters, in particular in the $\{111\}$ plane, which is presented subsequently.

\subsection{Vacancy clusters in $\{111\}$ and prismatic loops}
Experimental investigations indicate $\{111\}$ as one of the potential habit planes for prismatic dislocation loop nucleation from vacancy clustering~\citep{kuhlmann1960behavior,Cotterill1963}. It is hypothesized that when sufficiently large number of vacancies form a vacancy cluster, the planes of atoms above and below the cluster can collapse to form dislocation loops. In particular, studies indicate that dislocation loops with hexagonal symmetry are commonly observed and are energetically favorable~\citep{kuhlmann1960behavior}. However, the observed sizes of the loops are often $50$~\AA~diameter or larger, which are formed from vacancy clusters containing hundreds of vacancies. While loops of these sizes have been experimentally observed, the nucleation of loops is a very rapid process that is hard to observe experimentally. 

\begin{figure}[htbp]%
\centering%
\scalebox{0.5}{\includegraphics{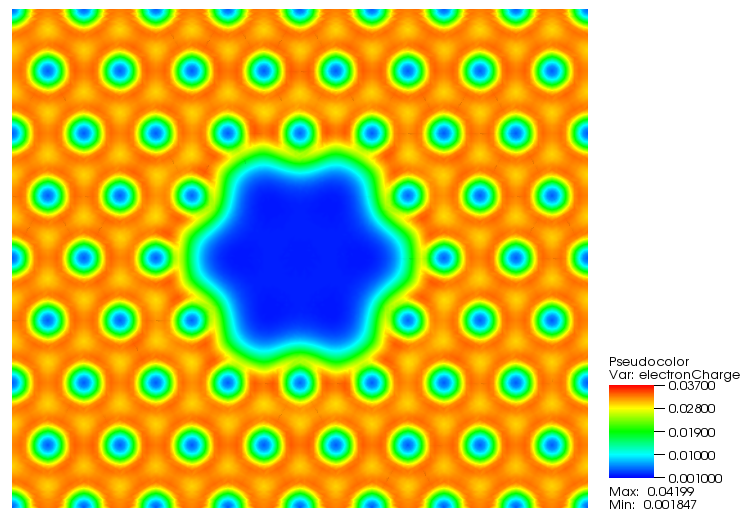}}%
\caption{\label{fig:hexVac} Electron-density contours of unrelaxed hexagonal $7$ vacancy cluster in $\{111\}$ plane.}%
\end{figure}

In this work, we study the energetics of $\{111\}$ hexagonal vacancy clusters to identify the nucleation size of vacancy prismatic loop in aluminum. As these are larger vacancy clusters, we employ a million atom simulation domain to accurately capture the electronic perturbations as well as the long-ranged elastic fields. We first begin with the study of the energetics of a 7-vacancy hexagonal cluster. Figure \ref{fig:hexVac} shows the electron-density contours of an unrelaxed 7-vacancy hexagonal cluster in $\{111\}$ plane. The relaxed binding energy of this vacancy cluster is computed to be $0.29~eV$. The positive binding energy suggests that this vacancy cluster is stable with respect to dissociation into monovacanices. However, the binding energy of this vacancy cluster is similar to the binding energy of quad-vacancy clusters on $\{111\}$ planes. Thus, this vacancy cluster can potentially dissociate into smaller-sized vacancy clusters---for instance a quad-vacancy cluster, a divacancy and a monovacancy. Further, the relaxed configuration of this hexagonal vacancy cluster did not represent a collapsed vacancy loop. This is contrary to a prior study~\citep{Gavini3}, which showed a collapsed prismatic vacancy loop from a 7-vacancy hexagonal vacancy cluster. This discrepancy is primarily due to the Thomas-Fermi-von Weizsacker family of kinetic energy functionals employed in the prior study, which do not have the correct linear response of a free-electron gas. The WGC kinetic energy functional employed in this work has the correct linear response of a free-electron gas, and benchmark studies have shown this kinetic energy functional to be in good agreement with Kohn-Sham DFT for a wide range of material properties in aluminum.  

\begin{figure}[htbp]%
\centering%
\scalebox{0.5}{\includegraphics{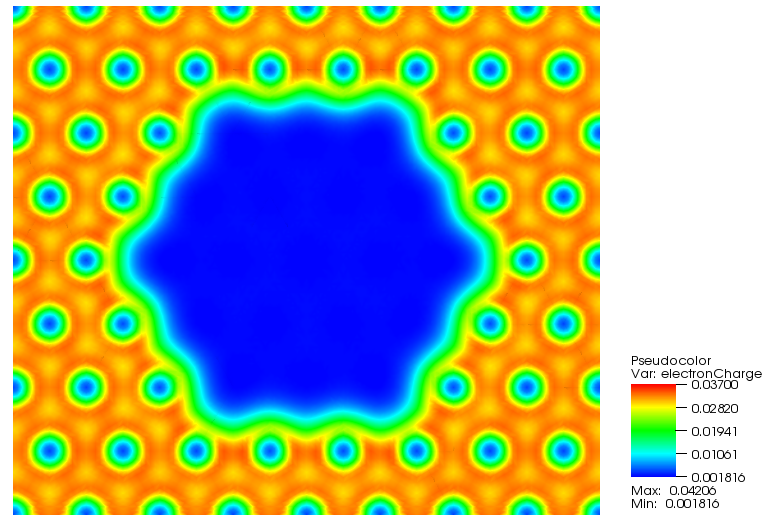}}%
\caption{\label{fig:hexVac2} Electron-density contours of unrelaxed hexagonal $19$ vacancy cluster in $\{111\}$ plane.}%
\end{figure}

\begin{figure}[htbp]
\centering
\scalebox{0.6}{\includegraphics{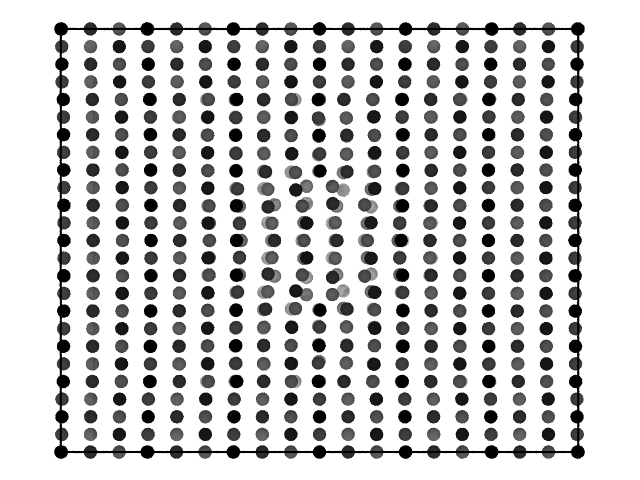}}%
\caption{\label{fig:loopCollapse} Atomic positions showing the collapse of vacancy cluster while viewing along $\left<112\right>$. The atoms are color coded in the gray scale with depth along the viewing direction.}%
\end{figure}

\begin{figure}[htbp]
\centering
\scalebox{0.8}{\includegraphics{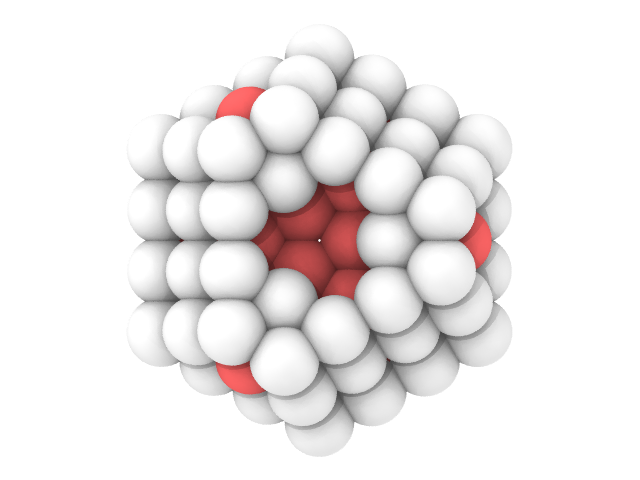}}%
\caption{\label{fig:coreStructure} The core of the vacancy prismatic loop identified using coordination analysis. The depicted atoms are those in the simulation domain that are not in an fcc coordination.}%
\end{figure}

We next study the energetics of the $\{111\}$ hexagonal vacancy cluster containing 19 vacancies. Figure \ref{fig:hexVac2} shows the electron-density contours of the unrelaxed hexagonal vacancy cluster. The relaxed binding energy of this hexagonal vacancy cluster is computed to be $10.02~eV$, which is a very large binding energy and significantly larger than that of the 7-vacancy cluster. The relaxed structure of this 19-vacancy hexagonal cluster resembles that of a collapsed prismatic loop. Figure~\ref{fig:loopCollapse} shows the relaxed atomic structure viewed along the $\left<112\right>$ direction, which shows the collapse of planes to form a vacancy prismatic loop. Figure~\ref{fig:coreStructure} shows the atoms in the simulation domain that do not have an face-centered-cubic (fcc) coordination which depicts the core of the prismatic dislocation loop. The Burgers vector, based on the displacement fields of the collapsed loop, is calculated to be ($\frac{1}{3}[111]$, $\frac{1}{18}[11\bar{2}]$, $\frac{1}{13}[1\bar{1}0]$). The major component of the collapse of the planes surrounding the vacancy loop is along $\left<111\right>$ direction, which is perpendicular to the loop. In addition, the planes also slip along the $\left<112\right>$ and $\left<110\right>$ directions. The slip along the $\left<112\right>$ direction is of particular importance. While the collapse along the $\left<111\right>$ direction creates a prismatic dislocation loop, it also encloses a stacking fault which is energetically unfavorable. The slip along $\left<112\right>$ eliminates the stacking fault, thus resulting in an unfaulted dislocation loop. The slip observed in the 19-vacancy loop is not sufficient to completely remove the stacking fault, but shows the tendency of the prismatic loop to unfault and we expect to observe complete unfaulting for larger-sized vacancy clusters. The present results suggest that vacancy clusters as small as 19 vacancies can collapse to form very stable prismatic dislocation loops, and represents the nucleation size of vacancy prismatic loops lying in the \{111\} planes.  

\section{Conclusion}\label{sec:Conclusion}
We conducted large-scale electronic structure calculations by employing quasi-continuum orbital-free DFT to investigate the energetics of vacancy clustering in aluminum. In particular, we used the WGC orbital-free kinetic energy functional in this study, whose accuracy has been ascertained for a wide range of properties in aluminum. The quasi-continuum reduction technique employed in this work has allowed electronic structure calculations using orbital-free DFT on multi-million atom simulation domains. Our cell-size study of binding energies of vacancies has shown that cell-sizes in excess of 2,000 atoms are required to obtain a converged value even for simple defects such as a divacancy. The binding energies of the divacancies and quad-vacancy clusters considered in this study are computed to be positive, suggesting the tendency of vacancies to coalesce to form vacancy clusters. Among the planar quad-vacancy clusters the one on $\{111\}$ had the highest binding energy, which is also a habit plane for the vacancy prismatic loops observed in experimental investigations. Investigations on hexagonal vacancy clusters of varying sizes showed positive binding energies. However, the 19-vacancy hexagonal vacancy cluster had significantly higher binding energy compared to the 7-vacancy hexagonal cluster. An analysis of the relaxed atomic positions of the 19-vacancy cluster revealed a collapsed vacancy prismatic dislocation loop. This study suggests that vacancy prismatic loops formed from vacancy clusters as small as 19 vacancies are stable and suggests that the nucleation sizes of these defects can be much smaller than the stable loops observed in experimental investigations that are often $50$~\AA\, or larger in diameter. 

While this study has shown that vacancy clustering is an energetically feasible mechanism that can result in the nucleation of vacancy prismatic loops, many outstanding questions remain and are worthwhile topics for future investigations. An interesting and important question is relative stability of non-planar vacancy clusters of various sizes in comparison to planar vacancy clusters, which has important bearing on the formation of stacking fault tetrahedra. Another key question relates to the interaction of these defects with dislocations that influences the macroscopic mechanical properties of materials. Further, while orbital-free DFT provides important insights into the mechanisms and qualitative trends in the energetics, conducting similar studies using Kohn-Sham DFT will provide quantitatively accurate energetics.       

\section*{Acknowledgments}
We thank Dr. Amuthan Arunkumar Ramabathian for assistance in the visualization of the atomic structure of the prismatic loop. We are grateful to the support of National Science Foundation (Grant number CMMI0927478) under the auspices of which this work was conducted. V.G. also gratefully acknowledges the hospitality of the Division of Engineering and Applied Sciences at the California Institute of Technology while preparing this manuscript. We also acknowledge the Advanced Research Computing at University of Michigan for providing the computing resources through the Flux computing platform.


\begin{thebibliography}{99}
\bibitem[Ackland et al. (1997)]{Ackland1997}
Ackland, G.J., Bacon, D.J., Calder, A.F., Harry, T., 1997. Computer simulation of point defect properties in dilute Fe-Cu alloy using a many-body interatomic potential. Phil. Mag. A 75, 713-732.

\bibitem[Arakawa et al.(2007)]{Arakawa2007}
Arakawa, K., Ono, K., Isshiki, M., Mimura, K., Uchikoshi, M., Mori, H., 2007. Observation of the one-dimensional diffusion of nanometer-sized dislocation loops. Science 318, 956-959.

\bibitem[Bacon et al.(1993)]{Bacon1993}
Bacon, D.J., Calder, A.F., Harder, J.M., Wooding, S.J., Computer simulations of low energy displacement in pure bcc and hcp metals, J. Nucl. Mater. 205, 52-58.

\bibitem[Bacon \& de la Rubia(1994)]{Bacon1994}
Bacon, D.J., de la Rubia, T.D., 1994. Molecular dynamics computer simulations of displacement cascades in metals. J. Nucl. Mater. 216, 275-290.

\bibitem[Barnes \& Mazey(1960)]{Barnes1960}
Barnes, R.S., Mazey, D.J., 1960. The nature of radiation-induced point defect clusters. Phil. Mag. 5, 1247-1253.

\bibitem[Blanc et al.(2002)]{Blanc}
Blanc, X., Le Bris, C., Lions, P.L., 2002. From molecular models to continuum mechanics. Arch. Rational Mech. Anal. 164, 341-381.

\bibitem[Carling \& Carter(2003)]{Carter-Al-Mg}
Carling, K.M., Carter, E.A., 2003. Orbital-free density functional theory calculations of the properties of Al, Mg and Al-Mg crystalline phases. Model. Simul. Mater. Sci. Eng. 11, 339-348.

\bibitem[Caturla et al.(2006)]{Caturla2006}
Caturla, M.J., Soneda, N., de la Rubia, T.D., Fluss, M., 2006. Kinetic Monte Carlo simulations applied to irradiated materials: The effect of cascade damage in defect nucleation and growth. J. Nucl. Mater. 351, 78-87.

\bibitem[Choly \& Kaxiras(2002)]{Choly2002}
Choly, N., Kaxiras, E., 2002. Kinetic energy functionals for non periodic systems. Solid State Commun. 121, 281-286.

\bibitem[Cotterill \& Segalla(1963)]{Cotterill1963}
Cotterill, R.M.J., Segalla, R.L., 1963. The effect of quenching history, quenching temperature and trace impurities on vacancy clusters in aluminium and gold. Phil. Mag. 8, 1105-1125.

\bibitem[Das et al.(2015)]{Das2015}
Das, S., Iyer, M., Gavini, V., 2015. An efficient real-space formulation of orbital-free density functional theory using finite-element discretization for Al, Mg and Al-Mg intermetallics. arXiv:1504.06368.

\bibitem[Eyre \& Bartlett (1965)]{eyre1965electron}
Eyre, B.L., and Bartlett, A.F., 1965. An electron microscope study of neutron irradiation damage in $\alpha$-iron. Phil. Mag. 12, 261-272.

\bibitem[Finnis(2003)]{Finnis}
Finnis, M., 2003. Interatomic forces in condensed matter, Oxford University Press, Oxford.

\bibitem[Gavini et al.(2007a)]{Gavini2}
Gavini, V., Bhattacharya, K., Ortiz, M., 2007a. Quasi-continuum orbital-free density-functional theory: A route to multi-million atom non-periodic DFT calculation. J. Mech. Phys. Solids 55, 697-718.

\bibitem[Gavini et al.(2007b)]{Gavini3}
Gavini, V., Bhattacharya, K., Ortiz, M., 2007b. Vacancy clustering and prismatic dislocation loop formation in aluminum. Phys. Rev. B 76, 180101(R).

\bibitem[Gavini et al.(2007c)]{Gavini1}
Gavini, V., Knap, J., Bhattacharya, K., Ortiz, M., 2007c. Non-periodic finite-element formulation of orbital-free density-functional theory. J. Mech. Phys. Solids 55, 669-696.

\bibitem[Gavini(2008)]{GaviniPRL2008}
Gavini, V., 2008. Role of macroscopic deformations in energetics of vacancies in aluminum. Phys. Rev. Lett. 101, 205503.


\bibitem[Gavini \& Liu (2011)]{GaviniJMPS2011}
Gavini, V., Liu, L., 2011. A homogenization analysis of the field theoretical approach to the quasi-continuum method. J. Mech. Phys. Solids 59, 1536-1551.


\bibitem[Gouldstone et al.(2000)]{Suresh2000}
Gouldstone, A., Koh, H.J., Zeng, K.Y., Giannakopoulos, A.E., Suresh, S., 2000. Discrete and continuous deformation during nanoindentation of thin films. Acta Mater. 48, 2277-2295.

\bibitem[Han et. al(2003)]{han2003interatomic}
Han, S., Zepeda-Ruiz, L.A., Ackland, G.J., Car, R., Srolovitz, D.J., 2003. Interatomic potential for vanadium suitable for radiation damage simulations. J. Appl. Phys. 93, 3328-3335.

\bibitem[Ho et al.(2007)]{Ho}
Ho, G., Ong, M.T., Caspersen, K.J., Carter, E.A., 2007. Energetics and kinetics of vacancy diffusion and aggregation in shocked aluminum via orbital-free density functional theory. Phys. Chem. Chem. Phys. 9, 4951.
 
\bibitem[Hung et al.(2010)]{Profess2}
Hung, L., Huang, C., Shin, I., Ho, G., Ligneres, V.L., Carter, E.A., 2010. Introducing PROFESS 2.0: A parallelized, fully linear scaling program for orbital-free density functional theory calculations. Comput. Phys. Comm. 181, 2208-2209.

\bibitem[Iyer \& Gavini(2011)]{Iyer2011field}
Iyer, M., Gavini, V., 2011. A field theoretic approach to the quasi-continuum method. J. Mech. Phys. Solids 59, 1506-1535.

\bibitem[Iyer et al.(2014)]{Iyer2014}
Iyer, M., Pollock, T.M., Gavini, V., 2014. Energetics and nucleation of point defects in aluminum under extreme tensile hydrostatic stresses. Phys. Rev. B 89, 014108 (2014).

\bibitem[Iyer et al.(2015)]{Iyer2015}
Iyer, M., Radhakrishnan, B.G., Gavini, V., 2015. Electronic-structure study of an edge dislocation in aluminum and the role of macroscopic deformations on its energetics. J. Mech. Phys. Solids 76, 260-275. 

\bibitem[Kiritani(1997)]{Kiritani1997}
Kiritani, M., 1997. Story of stacking fault tetrahedra. Mater. Chem. Phys. 50, 133-138.

\bibitem[Kiritani et al.(1999)]{Kiritani1999}
Kiritani, M., Satoy, Y., Kizuka, Y., Arakawa, K., Ogasawara, Y., Arai, S., Shimomura, Y., 1999. Anomalous production of vacancy clusters and the possibility of plastic deformation of crystalline metals without dislocations. Phil. Mag. 79, 797-804.

\bibitem[Knap \& Ortiz (2001)]{knap2001analysis}
Knap, J., Ortiz, M., 2001. An analysis of the quasicontinuum method. J. Mech. Phys. Solids 49, 1899-1923.

\bibitem[Kuhlmann et al. (1960)]{kuhlmann1960behavior}
Kuhlmann-Wilsdorf, D. and Wilsdorf, H.G.F., 1960. On the behavior of thermal vacancies in pure aluminum. J. App. Phys., 31, 516-525.

\bibitem[Lubarta et al.(2004)]{Lubarta2004}
Lubarda, V.A., Schneider, M.S., Kalantar, D.H., Remington, B.A., Meyers, M.A., 2004. Void growth by dislocation emission. Acta Mater. 52, 1397-1408.

\bibitem[Martin(2011)]{Martin}
Martin, R.M. 2011. Electronic structure: Basic theory and practical methods, Cambridge University Press, Cambridge.

\bibitem[Masters(1965)]{Masters1965}
Masters, B.C., 1965. Dislocation loops in irradiated iron. Philos. Mag. 11, 881-893.

\bibitem[Marian et al.(2002a)]{Marian2002PRL}
Marian, J., Wirth, B.D., Perlado, J.M., 2002a. Mechanism of formation and growth of $\left<100\right>$ interstitial loops in ferritic materials. Phys. Rev. Lett. 88, 255507.

\bibitem[Marian et al.(2002b)]{Marian2002PRB}
Marian, J., Wirth, B.D., Caro, A., Sadigh, B., Odette, G.R., Perlado, J.M., de la Rubia, T.D., 2002b. Dynamics of self-interstitial cluster migration in pure $\alpha$-Fe and Fe-Cu alloys. Phys. Rev. B 65, 144102.

\bibitem[Matsukawa \& Zinkle(2007)]{Zinkle2007}
Matsukawa, Y., Zinkle, S.J., 2007. One-Dimensional Fast Migration of Vacancy Clusters in Metals. Science 318, 959-962. 

\bibitem[Morishita et al.(2003)]{Morishita2003}
Morishita, K., Sugano, R., Wirth, B.D., 2003. MD and KMC modeling of the growth and shrinkage mechanisms of helium-vacancy clusters in Fe. J. Nucl. Mater. 323, 243-250.

\bibitem[Motamarri et al.(2012)]{Motamarri-OFDFT2012}
Motamarri, P., Iyer, M., Knap, J., Gavini, V., 2012. Higher-order adaptive finite-element methods for orbital-free density functional theory. J. Comp. Phys. 231, 6596-6621.

\bibitem[Motamarri et al.(2013)]{Motamarri-KSDFT2013}
Motamarri, P., Nowak, M.R., Leiter, K., Knap, J., Gavini, V., 2013. Higher-order adaptive finite-element methods for Kohn-Sham density functional theory. J. Comp. Phys. 253, 308-343.


\bibitem[Parr \& Yang(2003)]{Parr}
Parr, R., Yang, W., 2003. Density-functional theory of atoms and molecules, Oxford University Press, Oxford.

\bibitem[Perdew \& Zunger(1981)]{Perdew}
Perdew, J.P., Zunger, A., 1981. Self-interaction correction to density-functional approximations for many-electron systems. Phys. Rev. B 23, 5048.

\bibitem[Radhakrishnan \& Gavini(2010)]{GaviniPRB2010}
Radhakrishnan, B., Gavini, V., 2010. Effect of cell size on the energetics of vacancies in aluminum studied via orbital-free density functional theory. Phys. Rev. B 82, 094117.

\bibitem[Rice \& Zinkle(1998)]{Zinkle1998}
Rice, P.M., Zinkle, S.J., 1998. Temperature dependence of the radiation damage microstructure in V-4Cr-4Ti neutron irradiated to low dose. J. Nucl. Mater. 258-263, 1414-1419. 

\bibitem[Robinson (1994)]{robinson1994basic}
Robinson, M.T., 1994. Basic physics of radiation damage production. J. Nucl. Mater. 216, 1-28.

\bibitem[Singh et. al. (1997)]{singh1997radiation}
Singh, B. N., Foreman, A.J.E., and Trinkaus, H., 1997. Radiation hardening revisited: role of intracascade clustering. J. Nucl. Mater. 249, 103-115.

\bibitem[Soneda \& de la Rubia(1998)]{Rubia1998}
Soneda, N., de la Rubia, T.D., 1998. Defect production, annealing kinetics and damage evolution in $\alpha$-Fe: an atomic-scale computer simulation. Phil. Mag. A 78, 995-1019.  


\bibitem[Tadmor et al.(1996)]{Tadmor1996}
Tadmor, E.B., Ortiz, M., Phillips, R., 1996. Quasicontinuum analysis of defects in solids. Philos. Mag. A 73, 1529-1563.

\bibitem[Trachenko et. al (2001)]{trachenko2001atomistic}
Trachenko, K.O., Dove, M.T., Salje, E.K.H., 2001. Atomistic modeling of radiation damage in zircon. J. Phys.: Condens. Matter 13, 1947-1959.

\bibitem[Trinkaus et. al. (1997)]{trinkaus1997segregation}
 Trinkaus, H., Singh, B.N., and Foreman, A.J.E., 1997. Segregation of cascade induced interstitial loops at dislocations: possible effect on initiation of plastic deformation. J. Nucl. Mater. 251, 172-187.

\bibitem[Wang et al.(1999)]{Yan2}
Wang, Y.A., Govind, N., Carter, E.A., 1999. Orbital-free kinetic-energy density functionals with a density-dependent kernel. Phys. Rev. B 60, 16350.




\bibitem[Wirth et al.(1997)]{Wirth1997}
Wirth, B.D., Odette, G.R., Maroudas, D., Lucas, G.E., 1997. Energetics of formation and migration of self-interstitials and self-interstitial clusters in $\alpha$-iron. J. Nucl. Mater. 244, 185-194.

\bibitem[Wirth et al.(2000)]{Wirth2000}
Wirth, B.D., Odette, G.R., Maroudas, D., Lucas, G.E., 2000. Dislocation loop structure, energy and mobility of self-interstitial atom clusters in bcc iron. J. Nucl. Mater. 276, 33-40. 

\bibitem[Wirth(2007)]{Wirth2007}
Wirth, B.D., 2007. How does radiation damage materials? Science 318, 923-924.

\bibitem[Wu et al.(2007)]{Wu2007}
Wu, X.L., Li, B., Ma, E., 2007. Vacancy clusters in ultrafine grained Al by severe plastic deformation. Appl. Phys. Lett. 91, 141908. 

\bibitem[Zhou et. al (2004)]{zhou2004transferable}
Zhou, B., Wang, Y.A., and Carter, E.A., 2004. Transferable local pseudopotentials derived via inversion of the Kohn-Sham equations in a bulk environment. Phys. Rev. B 69, 125109.

\bibitem[Zinkle et. al (2011)]{Zinkle2011}
Zinkle, S.J., Ghoniem, N.M., 2011. Prospects for accelerated development of high performance structural materials. J. Nucl. Mater. 417, 2-8.


\end{thebibliography}
\end{document}